\documentclass[11pt,twoside]{article}

\usepackage{graphicx}
\usepackage{asp2006}
\usepackage{epsf}
\usepackage{psfig}
\usepackage{lscape}

\begin{document}
\title{ The transition from AGB to post-AGB evolution as observed by AKARI and
Spitzer}   
\author{D. Engels$^1$, P. Garc\'{i}a-Lario$^2$, F. Bunzel$^1$, D.A. Garc\'{i}a-Hern\'{a}ndez$^3$, J.V. Perea-Calder\'{o}n$^2$}   
\affil{$^1$ Hamburger Sternwarte, Universit\"at Hamburg, Germany \\
$^2$ European Space Astronomy Centre, ESAC, ESA, Madrid, Spain \\
$^3$ Instituto de Astrof\'{i}sica de Canarias, Tenerife, Spain \\
}    

\begin{abstract} 
  The AKARI and Spitzer satellites provided an unique opportunity to
  observe a variety of stars, which are considered as departing from
  the Asymptotic Giant Branch (AGB) and have started their post-AGB
  evolution recently. Most of these stars are absent optically and are
  bright in the mid-IR wavelength range. Spectra of close to 200
  objects have been obtained.  For all of them the $1-60\mu$m spectral
  energy distribution has been constructed using photometric data from
  various surveys. We report here on the results of Spitzer
  observations of 88 IRAS selected post-AGB candidates and discuss
  them in comparison to the results of the AKARI observations of
  post-AGB candidates reported elsewhere in these proceedings.  The
  dust compositions can be divided broadly in oxygen- and carbon-rich
  types, but a variety of intermediate types have been found. Among
  the oxygen-rich stars amorphous dust prevails, but a few sources
  show emission features from crystalline dust. The spectra from
  carbon-rich shells may be completely featureless, may show emission
  features from PAHs or a molecular absorption line from C$_2$H$_2$.
  We found also sources with a neon emission line at $12.8\mu$m. More
  than a third of all sources show a near-infrared excess at $\lambda
  < 5 \mu$m and almost all of them show evidence of C-rich dust in
  their shells. We postulate that the emerging post-AGB wind after the
  end of AGB evolution contains always carbon-rich dust irrespective
  of the chemistry of the former AGB star.

\end{abstract}



\section{The hidden phase of post-AGB evolution}
At the end of the stellar evolution on the Asymptotic Giant Branch
(AGB) stars loose copious amounts of mass, which build up a
circumstellar dust and gas shell hiding the star from optical view
almost completely. Stars departing from the AGB and evolving towards
the Planetary Nebula (PN) phase are therefore difficult to observe
optically. It was found that a number of Proto-Planetary Nebulae (cf.
in CRL 2688; \citeauthor{sahai98} \citeyear{sahai98}) show high
velocity bipolar outflows which are connected to a fast,
axially-symmetric wind, which is taking the place of the much slower,
spherically-symmetric wind operating on the AGB. The physical
mechanism responsible for the change of the spherically-symmetric to
an axially-symmetric, or in some cases point-symmetric wind is
strongly debated. Observations of masers in transition objects often
reveal that this morphological change takes place at a very early
stage in the post-AGB phase (\citeauthor{sahai99} \citeyear{sahai99};
\citeauthor{zijlstra01} \citeyear{zijlstra01}), while the star is still
heavily obscured in the optical range.

Non-variable OH/IR stars \citep{engels02} and IRAS selected infrared
sources with extreme red colors \citep{suarez06} are
candidates for such hidden post-AGB stars. Their study has made
progress only in the last decade due to the improved observation
capabilities in the infrared at $\lambda > 5 \mu$m by space-based
observatories. In the mid-infrared the emission emerges from the 
circumstellar envelopes (CSE)
and their gas and dust composition has to be used to infer on the
evolutionary state of the underlying star and the mass loss process. 

The spectroscopic observations with ISO showed that strong changes
occur in the infrared SEDs during total obscuration \citep{garcia03}.
In the case of the C-rich AGB stars the
molecular C$_2$H$_2$ absorption and the amorphous SiC emission feature at
11.3$\mu$m suddenly disappear and become substituted by a broad
plateau of emission from 11 to 15$\mu$m due to hydrogenated PAHs.
These are later replaced by de-hydrogenated,
narrow PAH features at 3.3, 6.2, 7.7, 8.6 and 11.3$\mu$m, which are
also observed in more evolved C-rich PNe. In  O-rich AGB stars the
strong silicate absorption features at 9.7 and 18$\mu$m  disappear and 
are replaced by several prominent
crystalline silicate emission features in the $10-40\mu$m wavelength
range.  A mixed chemistry is found also in a few sources, but it
is not clear whether it is associated to late thermal pulses at the
end of the AGB phase and/or to the preservation of O-rich material in
long-lived circumstellar disks.  Globally considered, there seems to
be a continuous evolution from an amorphous (aliphatic) to crystalline
(aromatic) organization of molecules in the dust grains both in the
C-rich and the O-rich sequence, which is still unexplained \citep{garcia06}.

\section{AKARI and Spitzer observations of hidden post-AGB stars}
The AKARI satellite \citep{murakami07} and the Spitzer Space Telescope
\citep{werner04} offered the possibility to extend the ISO
observations to larger and better selected samples of hidden post-AGB
stars. Observations between 2 and 26$\mu$m were possible with the
Infrared Camera (IRC) \citep{onaka07} on board of AKARI, and in the
range $5 < \lambda < 37\mu$m with the InfraRed Spectrograph (IRS)
\citep{houck04} on board of Spitzer. A first sample studied consisted
of obscured OH/IR sources with associated radio continuum emission.
The Spitzer spectra allowed a re-classification of the sources in AGB
stars and post-AGB stars close to the formation of PNs
\citep{hernandez07}. The new samples observed, consisted of extremely
red IRAS sources from the GLMP catalog \citep{garcia92}, and of OH/IR
stars selected on the base of their appearance in the Spitzer GLIMPSE
survey. The 2MASS and GLIMPSE surveys were used to identify OH/IR
stars with near-infrared excesses indicative for a post-AGB nature of
these sources \citep{engels07}.

\begin{figure}[!ht]
\begin{center}
 {\includegraphics*[width=7cm,angle=-90]{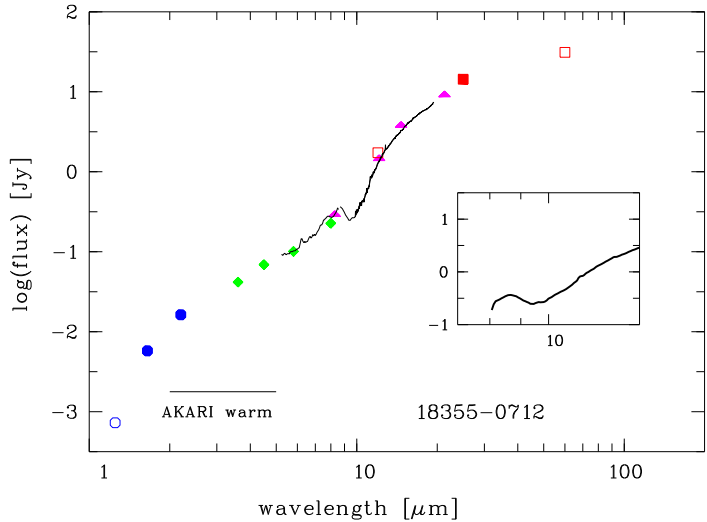}}
\end{center}
\caption{ Spitzer $5<\lambda<20 \mu$m spectrum of IRAS\,18355-0712
(black) superposed on the $1-60\mu$m spectral energy distribution
constructed from photometric data taken from the 2MASS (blue dots),
GLIMPSE (green diamonds), MSX (magenta triangles), and IRAS (red
squares) surveys. Open symbols represent low-quality data.  
This object is an example for a ``non-variable
OH/IR star'' classified as post-AGB star with red mid-IR continuum, a
weak silicate feature and excess emission in the near-infrared.
}\label{fig:sp18355}
\end{figure}

\subsection{AKARI observations of OH/IR stars and extreme carbon stars}
The SEDs of obscured variable OH/IR stars peak in the wavelength range
$5-30\mu$m and show a strong 10$\mu$m and a weaker 18$\mu$m absorption
feature. These SEDs can be modeled in detail using cold dust opacity
functions of amorphous silicates \citep{suh02}. This is confirmed by
the results we obtained from the modeling of the AKARI spectra of the
infrared sources classified as AGB stars (Bunzel et al., these
proceedings). The carbon-rich cousins of OH/IR stars are the 'extreme
carbon stars' (extreme in terms of infrared color). The dust features
seen in their SEDs are usually weak, but they often show a molecular
absorption line at $13.7 \mu$m attributed to C$_2$H$_2$. The extreme
carbon stars are rarer than OH/IR stars and harder to classify because
of the lack of prominent dust features and radio maser emission.
Before AKARI, the most extreme carbon stars were studied by
\citet{volk00}, who modeled the SEDs successfully with amorphous
carbon dust. They found the evolutionary status compatible with the
end phase of AGB evolution. The extreme carbon stars, we identified among
the infrared sources observed with AKARI, are the reddest found so far.
The spectra of all of them (except IRAS\,15408-5657) could be modeled
with amorphous carbon dust ($\eta_C>80\%$), with minor contributions
of SiC and silicates. Because of a low IRAS variability index we
suspect that part of them could have started post-AGB evolution
already (Garc\'{i}a-Hern\'{a}ndez et al., these proceedings).

IRAS\,15408-5657 is a peculiar source, in the sense that its silicate
absorption features are too weak for its red continuum. The model SED
required a mixture of carbon and silicate dust in almost equal parts
to obtain the appropriate strength of the silicate band. Its low IRAS
variability index makes it a post-AGB candidate. It is unlikely that
both dust species spatially coexist, because the underabundant atomic
species (C or O) should be locked in CO, and would not be available
for dust formation \citep{ivezic95}. Thus, the mixed chemistry may
indicate the presence of two shells, an inner shell with C-rich dust
and an outer one with O-rich dust.

\begin{figure}[!ht]
\begin{center}
 {\includegraphics*[width=5cm,angle=-90]{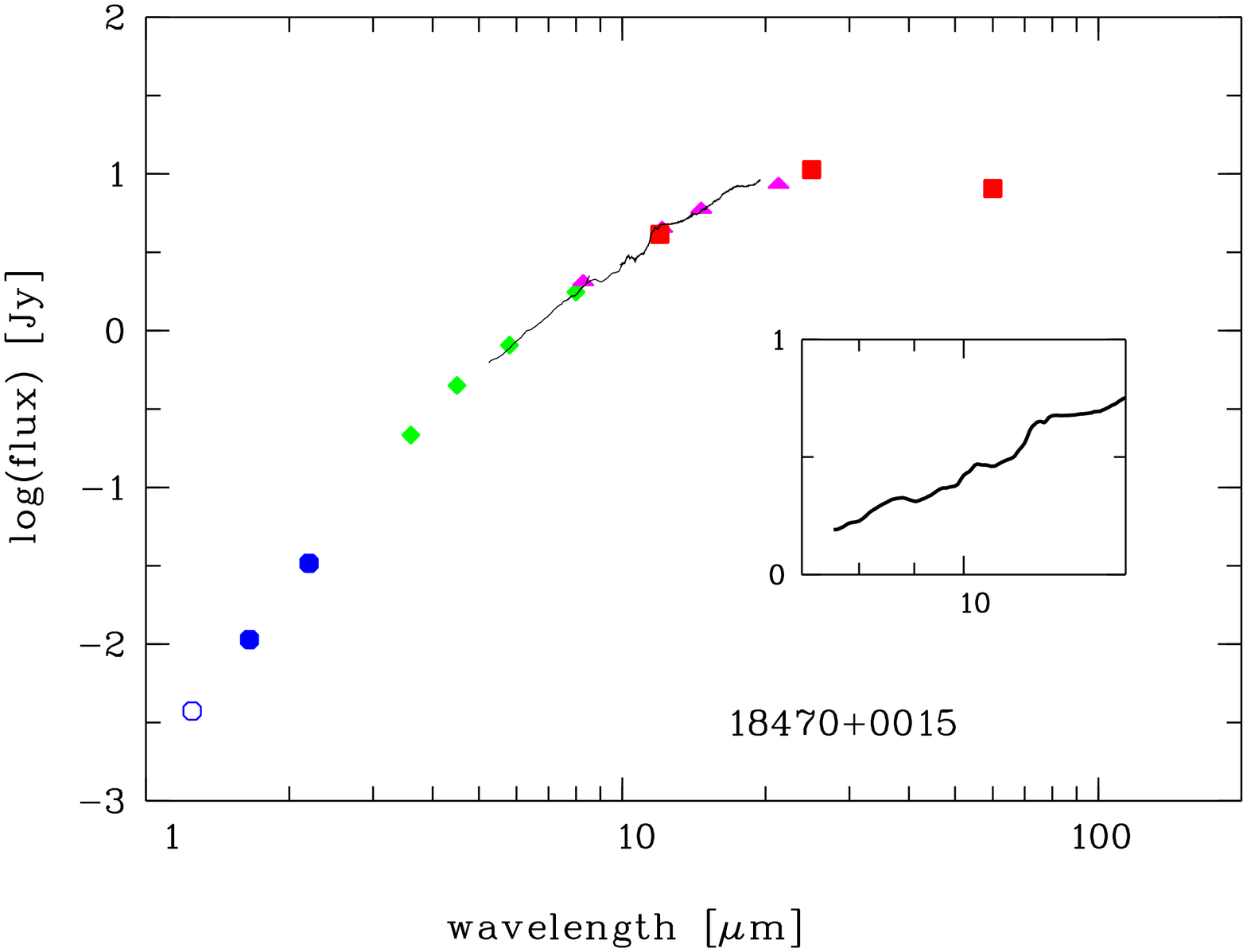}}
 {\includegraphics*[width=5cm,angle=-90]{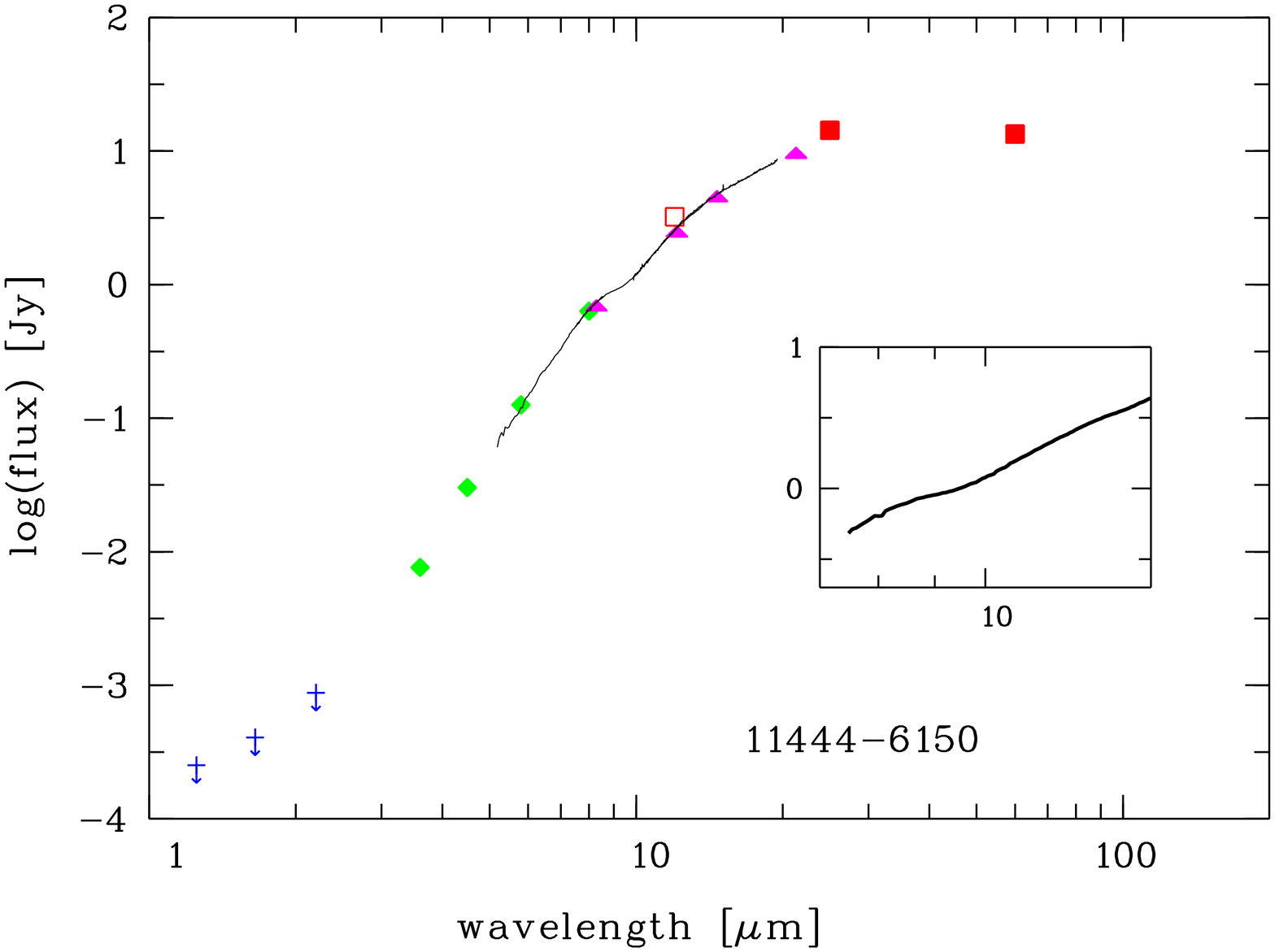}}
\end{center}
\caption{Spitzer $5<\lambda<20 \mu$m spectra of IRAS\,18470+0015 (a)
and IRAS\,11444-6150 (b). Symbols as in Fig. \ref{fig:sp18355}. Arrows
denote upper limits. IRAS\,18470+0015 shows weak features of
crystalline silicate dust. All three objects of this type have OH masers.
IRAS\,11444-6150 is a carbon-rich post-AGB star with a spectrum classified
as 'featureless'.
}\label{fig:sp18470}
\end{figure}

For several sources with silicate absorption features observed by
Bunzel et al. with AKARI we had indications for their post-AGB nature
beforehand. Either due to the presence of bipolar high-velocity
outflows traced by the H$_2$O masers (IRAS\,19134+2131,
\citeauthor{imai04} \citeyear{imai04}; OH\,31.0+0.0 = W43A,
\citeauthor{imai02} \citeyear{imai02}), or due to the presence of a
near-infrared excess. As for IRAS\,15408-5657 the SEDs of these sources
could not be modeled by pure amorphous silicate dust, but required a
model, where the inner carbon-rich shell is viewed through an outer
shell containing 20-40\% silicate dust. The results for these post-AGB
stars and for IRAS\,15408-5657 indicate that for oxygen-rich AGB stars
the departure from the AGB marks also a change in dust chemistry.
Carbon-rich dust forms in the inner shell, while the silicate-rich
dust shell formed on the AGB expands outwards.


\subsection{Spitzer observations of the GLMP sample of post-AGB stars}
\begin{figure}[!ht]
\begin{center}
 {\includegraphics*[width=7cm,angle=-90]{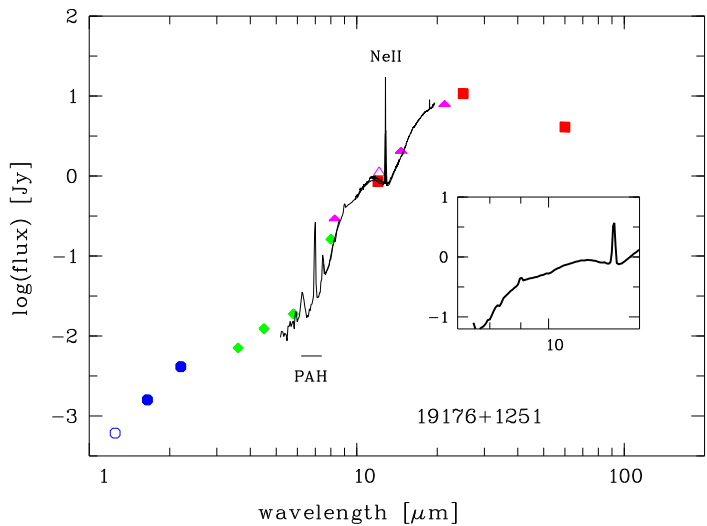}}
\end{center}
\caption{Spitzer $5<\lambda<20 \mu$m spectrum of IRAS\,19176+1251.
Symbols as in Fig. \ref{fig:sp18355}. Strong PAH features  were
found only among objects showing a near-infrared excess at $\lambda < 5\mu$m.
}\label{fig:sp19176}
\end{figure}

A preliminary evaluation of the Spitzer spectra in the
$5-20\mu$m range and the $1-60\mu$m
SEDs of 88 IRAS sources from the GLMP catalog confirm the AKARI
results and show that the mid-IR spectra are even more diversified
than expected. Judged from the IRAS variability index almost all these
sources have a chance of $<50$\% to be variable and are therefore
currently post-AGB stars. Based on the spectra and the $1-60\mu$m SEDs
they can be divided into several groups:
\begin{itemize}
\item About 45\% have strong silicate absorption
features and are heavily obscured in the near-infrared. These are
former O-rich AGB stars, where the remnant AGB shell dominate the
mid-IR spectra.  
\item Another 15\% show the combination of a very red
continuum longward of $\lambda=5\mu$m and a weak silicate absorption.
About two-third of them have a near-infrared excess at $\lambda<5\mu$m
as it is exemplified in IRAS\,18355$-$0712 (Fig. \ref{fig:sp18355}).
These sources are similar to the AKARI observed post-AGB stars, which
required a mixed chemistry to model their SEDs.
\item Three sources show evidence for the presence of crystalline 
silicate dust judged from sharp absorption features in the $10\mu$m region. 
All of them are OH/IR stars with a relatively blue continuum 
(see Fig. \ref{fig:sp18470}a). 
\item Another 20\% show featureless spectra (see IRAS\,11444-6150 in Fig.
  \ref{fig:sp18470}b), spectra with C$_2$H$_2$ absorption at
  13.7$\mu$m or weak carbon dust features. Almost half of them show a
  near-infrared excess as in IRAS\,18355$-$0712. These objects are
  probably post-AGB carbon stars with varying degrees of optical
  depths of their remnant AGB shells and where the objects with a
  near-infrared excess are the more evolved.
\item The remaining objects have (in part strong) near-infrared
excesses and show a variety of carbon dust features in their spectra.
An example is IRAS\,19176+1251, which shows strong PAH features and in
addition a Ne{\sc{ii}} 12.8$\mu$m emission line, coming probably from
material shocked by the stellar wind (Fig. \ref{fig:sp19176}). These
objects are considered as the most advanced in their post-AGB
evolution.
\end{itemize}

\section{Conclusions}
Post-AGB evolution starts when almost the complete stellar envelope
has been lost by the stellar wind, and the stars are still hidden by
their circumstellar envelope. Such stars do not show the AGB typical
long-period variability anymore. The AKARI and Spitzer spectra of such
hidden post-AGB stars indicate that the inner part of the CSEs
contains carbon-rich dust formed in the post-AGB wind irrespective of
the chemistry of the star on the AGB. The silicate absorption features
seen in many of the sources may originate from the outer CSE, which is
composed mainly by dust of the remnant AGB shell. These conclusions
can be probed in those stars, where the remnant AGB shell has been
diluted far enough, that observations of the warm dust near the star
are possible.  Spectroscopy in the $2-5\mu$m range using the IRC
during the AKARI warm phase will therefore be made, to search for
C-rich matter in the inner dust shell of those post-AGB stars showing
the $10\mu$m silicate absorption and a strong near-infrared excess.

\acknowledgements 
This research is based on observations with AKARI, a
JAXA project with the participation of ESA, and on observations with
Spitzer, a NASA's Great Observatories Program. D. Engels acknowledges travel
support by the conference organizers.


\vspace{0.5cm}
{\bf Question by T. Onaka:} Is there any signature of C-rich material in the
NIR spectra of mixed chemistry objects?

{\bf Answer:} Not yet. Only with GLIMPSE covering the $3 < \lambda <
8\mu$m wavelength range it was possible to verify that the 2MASS
counterparts found for hidden post-AGB stars at $< 3\mu$m are not field
stars. Objects with confirmed near-infrared excess are currently
observed by AKARI between 2 and $5\mu$m.

{\bf Question by T. Onaka:} Do silicate dust grains have to have lower tem-
peratures than carbonaceous dust in the outer shell?

{\bf Answer:} This cannot be answered on the base of our DUSTY-based mod-
els.
\end{document}